\documentclass[runningheads]{llncs}
\usepackage[T1]{fontenc}
\usepackage{graphicx}
\usepackage{amsmath,amssymb,amsfonts}
\usepackage{amstext}
\usepackage{url}

\usepackage{algorithmic}

\usepackage{amscd,mathrsfs}
\usepackage{multirow}

\begin{document}
\title{Compact Artificial Neural Network Models for Predicting Protein Residue - RNA Base Binding}
\titlerunning{Compact ANN Models for Predicting Protein-RNA Binding}
%
\author{Stanislav Selitskiy \orcidID{0000-0003-1758-0171}}
\authorrunning{S. Selitskiy}
%
\institute{University of Bedfordshire, Park Square, Luton, LU1 3JU, UK\\
\email{stanislav.selitskiy@study.beds.ac.uk}
}

\maketitle              
\begin{abstract}
Large Artificial Neural Network (ANN) models have demonstrated success in various domains, including general text and image generation, drug discovery, and protein-RNA (ribonucleic acid) binding tasks. However, these models typically demand substantial computational resources, time, and data for effective training. Given that such extensive resources are often inaccessible to many researchers and that life sciences data sets are frequently limited, we investigated whether small ANN models could achieve acceptable accuracy in protein-RNA prediction. We experimented with shallow feed-forward ANNs comprising two hidden layers and various non-linearities. These models did not utilize explicit structural information; instead, a sliding window approach was employed to implicitly consider the context of neighboring residues and bases. We explored different training techniques to address the issue of highly unbalanced data. Among the seven most popular non-linearities for feed-forward ANNs, only three—Rectified Linear Unit (ReLU), Gated Linear Unit (GLU), and Hyperbolic Tangent (Tanh)—yielded converging models. Common re-balancing techniques, such as under- and over-sampling of training sets, proved ineffective, whereas increasing the volume of training data and using model ensembles significantly improved performance. The optimal context window size, balancing both false negative and false positive errors, was found to be approximately 30 residues and bases. Our findings indicate that high-accuracy protein-RNA binding prediction is achievable using computing hardware accessible to most educational and research institutions.

\end{abstract}
\section{Introduction}

Recent advancements in molecular biology have revealed the intricate and multifaceted roles of RNA, sparking significant interest in methods for predicting RNA-protein interactions and previously dismissed as "junk" sequences, large non-coding regions of RNA are now understood to be crucial in the actualization, regulation, and fine-tuning of genetic expression and evolution \cite{Li2019}. These regions, though not directly involved in the transcription process of converting DNA to RNA to protein, play essential roles in various regulatory mechanisms.

A comprehensive understanding of RNA functions, including DNA transcription, splicing mechanisms, interactions with nuclear pore proteins, nucleo-cytoplasmic transport, cytoplasmic motor protein functions, RNA localization, translation, post-translational modifications, and RNA folding stability, necessitates extensive research into RNA-protein binding. Additionally, RNA's role in encoding small regulatory peptides and non-protein-binding regulatory functions, such as low-affinity strand binding, further underscores the importance of studying these interactions on a larger scale than previously considered \cite{DiLiegro2014,nikitin2023non}.

To address these needs, researchers have developed various experimental techniques to identify RNA-protein interactions. These methods fall into two categories: RNA-centric (identifying proteins that bind to specific RNA sequences) and protein-centric (identifying RNA sequences that bind to specific proteins). While each method offers unique strengths for targeted studies, they also present limitations when applied to broader research contexts \cite{schmeing2009recent}. Despite advancements in high-throughput sequencing technologies for some proteins, the experimental validation of RNA-protein binding remains labor-intensive and limited in throughput \cite{licatalosi2008hits}.

In response to these challenges, computational machine learning (ML) approaches have been developed to facilitate the targeted experimental verification of promising RNA-protein interactions, offering a scalable and efficient alternative to traditional experimental methods.

\subsection{Existing Literature on Computational Protein-RNA Interaction Modeling}

The prediction of RNA-protein binding sites has emerged as a vibrant area of research, leveraging various types of input data, including sequencing information, secondary and tertiary structures, and physicochemical properties. Accurate and interpretable models for RNA-protein interactions necessitate a three-dimensional spatial understanding of protein structures, employing analytical approaches to identify structural similarities. Techniques such as template-based clustering \cite{zheng2016template}, energy function methods \cite{zhao2011structure}, electrochemical properties for homologous proteins (e.g., DR\_bind1) \cite{chen2014identifying}, and score-based methods like OPRA \cite{perez2010optimal} and KYG \cite{kim2006amino} have been pivotal in this domain. Despite their initial popularity in the early 2000s, these methods continue to evolve and remain relevant \cite{zheng2016template}.

However, obtaining the spatial structures of proteins and RNA is a labor-intensive experimental process, with known structures available for only a small subset of possible protein sequences \cite{thompson2020advances}. This limitation has driven extensive research into computational protein folding, aiming to predict 3D structures from sequence data alone \cite{dill2012protein}. Early combinatorial approaches \cite{huang2013novel} and recent advancements in large deep learning (DL) models, such as AlphaFold \cite{Jumper2021} and subsequent iterations, have achieved remarkable accuracy in solving the protein folding problem \cite{zhang2022few}. Nevertheless, these models demand significant computational resources, presenting a challenge for resource-constrained laboratory settings.

To mitigate the complexity of high-dimensional data, mixed 3D-2D models have been proposed. For instance, models integrating 2D RNA and 3D protein interactions using score matrices have shown promise \cite{Xie2020}. Further dimensionality reduction has led to the development of 1D or 1D+ models, often utilizing machine learning (ML) to capture subtle sequence-structure relationships that escape traditional analytical methods. Early shallow artificial neural networks (ANNs) \cite{jeong2004neural}, Naive Bayes classifiers like RNABindR \cite{terribilini2007rnabindr}, support vector machine (SVM) models such as BindN/BindN+ \cite{wang2006bindn}, Pprint \cite{kumar2008prediction}, PRINTR \cite{wang2008printr}, PiRaNhA \cite{murakami2010piranha}, and decision tree or random forest methods like NAPS \cite{carson2010naps} and PRBR \cite{ma2011prediction}, have all contributed significantly to this field. Hybrid SVM and evolutionary algorithms, such as RISP \cite{tong2008risp}, also play a crucial role. A more detailed comparison of these methods can be found in \cite{puton2012computational}. Despite originating in the early 2000s, these approaches are still actively researched, exemplified by recent work like the scoring SVM model of Kashiwagi et al. \cite{Kashiwagi2021}.

In recent years, the advent of large-scale DL ANNs with numerous parameters has become increasingly popular for molecular ML research \cite{schutt2018schnet}. Innovative ANN architectures that consider relationships between data entities, such as graph neural networks (GNNs) \cite{gasteiger2020directional} and physics-informed neural networks (PINNs) \cite{haghighatlari2022newtonnet}, have shown great potential. These models predict the energy state of entire molecular structures based on atomic descriptors and their spatial configurations \cite{pellegrini2023panna}. GNNs have been particularly useful for RNA \cite{joshi2023multi} and protein \cite{strokach2020fast} design, while attention mechanisms have enhanced the learning of molecular structures \cite{le2022representation} and protein-protein interactions \cite{asim2022adh}. Additionally, reinforcement learning (RL) has been applied to the design of protein and other biomolecular structures \cite{aderinwale2022rl}.

The goal of this study and future work is to leverage recent ANN methods within "Green AI" frameworks that emphasize small resource footprints, making advanced computational techniques accessible to researchers without extensive computational resources. This paper explores relatively shallow yet wide and straightforward feed-forward ANN models with limited non-linearity complexity, exploring their applicability to the RNA-protein binding problem.

\section{Materials and Methods}

\subsection{Data Set}

We utilized a data set of protein and RNA segments, indexed by binding positions, as described in \cite{Sato2021Oct} and constructed by the authors of \cite{Kashiwagi2021}. This data set is organized into an ``input" folder containing various file types: ``.fa" files for protein residues and RNA bases in alphabetical encoding, ``.ss" files for structural information encoded in ASCII symbols, and ``.dat" files for binding information as pairs of numeric positions of residues and bases in the corresponding protein and RNA segment ``.fa" files. The file correspondence is indicated by their names and listed in the training and test set index files ``train.lst" and ``test.lst". For example, the first line in the training set index might read: ``input/1n8rU.fa input/1n8rU.ss input/1n8rA.fa input/1n8rA.ss input/1n8rUA.dat". However, some index entries pointed to non-existent data files, necessitating a cleanup. The corrected index files, along with our source code, are available at \url{https://github.com/Selitskiy/Protein}.

\begin{figure}
\includegraphics[width=0.99\linewidth]{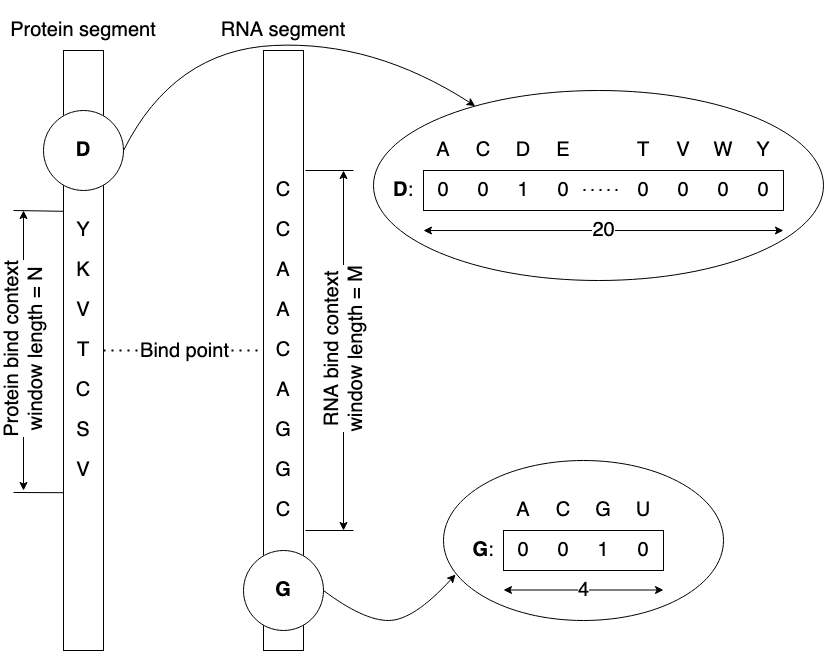}
\caption{Protein bind context window residues and RNA bind context window bases encoding schema.\label{fig.he}}
\end{figure}   

Each alphabetic representation of protein residues and RNA bases was converted into a ``one-hot-encoded" variable, as illustrated in Figure~\ref{fig.he}. For protein residues, a $20$-position variable was used, where the position corresponding to the residue's alphabetical order was set to $1$, while the other positions were set to $0$. The entire binding context variable is composed of an $N$-tuple of ``one-hot-encoded" protein residues in the context window centered around the binding point and an M-tuple of one-hot-encoded RNA bases, resulting in an overall length of $K=N*20+M*4$ informative positions. For practical encoding/decoding purposes, a sparse $26$-character English alphabet ASCII order was used with $6$ positions always being $0$. A total of $196,862$ binding examples were extracted from the data. If a binding position was closer than the context window, extra non-existent residues or bases were encoded by $0$ in all $20$ or $4$ one-hot-encoded positions, respectively, ensuring a consistent number of binding examples regardless of the context window lengths.

Non-binding examples were collected from the corresponding protein and RNA ``.fa" files. Segments outside the context windows around binding positions were split into non-overlapping fragments of the context window length and pairwise matched. Each protein segment without a binding point from a given ``.fa" file was joined with all RNA segments without a binding point from the corresponding ``*.fa" file. The number of non-binding examples thus depended on the context window length. For window lengths of $15$ residues/bases, the number of non-binding examples totaled $32,186,879$, and for lengths of $27$ residues/bases, the number was $18,607,564$.

\subsection{Machine Learning Models and Hardware}

Deep learning models, particularly large language models, graph ANNs, and physics- or chemistry-informed ANNs, have proven to be powerful tools for predicting protein-RNA interactions. However, the training of these models demands substantial resources, including data, time, financial investment, power, and a significant carbon footprint. Consequently, these models often remain proprietary, limiting their accessibility to the broader research community for use, development, and algorithm verification. This research explores whether smaller, shallow, but wide ANNs, which employ an implicit structure-preservation context window mechanism, can achieve sufficient accuracy while requiring considerably fewer resources for training.

The previous subsection details the preparation of input data, which includes the neighbouring context around the binding position. Our experimental ANN architecture is inspired by the concept of ANNs as universal function approximators based on the A. Kolmogorov and V. Arnold solution, Formula~\ref{eq:1}.
 
In this case, the function $f: \mathcal{X}\subset \mathbb{R}^K \mapsto \mathbb{R}$ which transforms $K$-dimensional input (representing the context window of protein residue-RNA base potential binding) into a single-dimensional softmax weight indicating actual binding.

\begin{equation}
\label{eq:1}
f(\textbf{x}) = f(x_1, \dots , x_m) = \sum_{q=0}^{2m} \Phi_q (\sum_{p=1}^{m} \phi_{qp}(x_p))
\end{equation}

where $\Phi_q$ and $\phi_{qp}$ are continuous $\mathbb{R} \mapsto \mathbb{R}$ functions \cite{kolmogorov1961representation}.

In the context of ANNs, the Kolmogorov-Arnold superposition theorem can be interpreted as a representation of a 2-layer ANN, where inputs to the inner functions $\phi_{qp}$ serve as dimension-specific nonlinearities within the perceptions. Such architecture allows using dynamic variability of activation functions to save on the model parameters number, compared to DL architectures.

\begin{figure}
\includegraphics[width=0.99\linewidth]{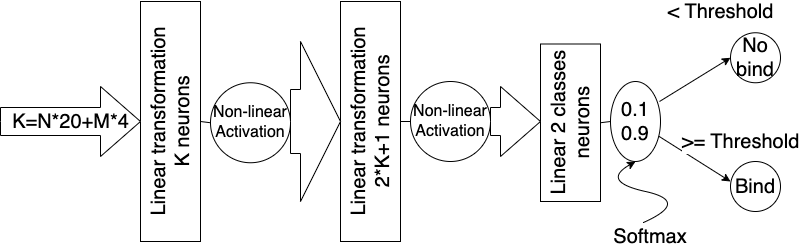}
\caption{High-level ANN schema used for experiments with different activation functions.\label{fig.ann}}
\end{figure}

We experimented with this ANN architecture, testing various activation functions to find a configuration that allows for efficient model conversion during training while maintaining high accuracy, Figure~\ref{fig.ann}. Theoretically, these activation functions can differ not only between layers but also between individual neurons. For this study, we focused on identifying a single class of activation functions suitable for all layers and neurons.

The experiments were performed on a system running Linux (Ubuntu 20.04.3 LTS), equipped with a Quadro RTX 8000 GPU (passive cooling, with 48GB GDDR6 memory), X299 chipset motherboard, 256GB DDR4 RAM, and an i9-10900X CPU. MATLAB 2022b with the Deep Learning Toolbox was used for the experiments. The ANNs were trained using the ``adam" learning algorithm with an initial learning rate of $0.01$. The mini-batch size was set to $524,288$, with a training duration of $50$ epochs, except for the Tanh ANN, which was trained for $150$ epochs due to reasons discussed later.

\subsection{Control Parameters}

We investigated the behaviour of the proposed ANN architecture and training methods by varying the following control parameters: activation function, training re-sampling algorithms, re-training algorithms, protein-RNA interaction window size, ensemble voting algorithms, and the presence of Dropout layers.

Initially, we experimented with seven ANN architectures using different activation functions under default computational settings to identify the most promising models. These activation functions included a simple linear regression ANN without non-linearities, as well as logistic sigmoid (Sigmoid), hyperbolic tangent (Tanh), rectified linear unit (ReLU), radial basis functions (RBF), KGate non-linearity \cite{Selitskiy2022}, and custom transformer-type non-linear activation \cite{Selitskiy2023Batch}.

Due to the non-balanced nature of the training set and hardware limitations, we employed re-balancing and re-training techniques. We created an over-sampled training set by duplicating $196,862$ bind examples $25$ times, resulting in $4,921,550$ examples paired with an equivalent number of no-bind examples. This no-bind set was split into two parts, and each model was re-trained with the same bind and the two no-bind subsets, creating an effective bind-to-no-bind ratio of $\frac{25}{50}$. For under-sampling, we randomly selected an equivalent number of bind and no-bind examples, achieving a $\frac{1}{2}$ ratio.

To handle the large data set that exceeded hardware capacity, we used successive re-training on randomly selected subsets. Since over- and under-sampling performed poorly in reducing false positive errors, we trained models with the maximum possible non-binding examples fitting into memory, resulting in a $\frac{1}{50}$ bind-to-no-bind ratio ($196,862$ bind and $9,843,100$ no-bind examples). The number of training epochs per re-train was scaled down to maintain the same overall epochs. For example, a single $50$-epoch run was divided into five 10-epoch re-trains. We experimented with binding context windows of $15$ and $27$ neighbouring amino residues and RNA bases around the bind position.

Given the significant influence of the context window on accuracy, various window lengths were tested for the ReLU model. A soft limit on the number of non-bind test examples was enforced to match the test set size across different context window lengths.

To reduce false positive errors, we experimented with homogeneous ensembles of the three convergent models using consensus and majority voting schemes. Based on false positive and negative error curves, a $27$-length context window was chosen for subsequent experiments.

We also evaluated the benefits of ensemble architectures, finding minor improvements for ReLU, but greater benefits for KGate with consensus voting and Tanh with majority voting. This led us to test heterogeneous ensembles of different ANN types, both pair-wise and all three together.

Finally, we compared models with and without Dropout layers to assess their impact on model regularization, similar to ensemble methods.

\subsection{Accuracy Metrics}
Traditional accuracy metrics such as Accuracy itself, Precision, Recall, Specificity (as a part of AUC calculation), $F1$ score \cite{davis2006relationship} were used to evaluate ANN architectures in this study.

The only worthy of detailed description details are descriptions of what are considered True and False Positives ($TP$, $FP$) and Negatives ($TN$, $FN$). Let's define $P$ as a set of observation $\forall o_i \in O$ classifications $c(o_i)$ as a protein-RNA bind position made on the condition of the ANN's softmax weight for the bind class $sm_1(o_i)$ greater or equal of the trusted threshold $Th$: $P = \{C:sm_1(o_i) \geq Th \}$, $O = B \cup NB$ is a set of all test observations composed of the subset of bind examples $B$ and no-bind examples $NB$. True Positives is a subset of $P$ which intersects with bind observations $TP = P \cap B$, and a number of True Positives $N_{TP} = |TP|$ is a power of the such set. Respectively, $N_{FP} = |P \cap NB|$, $N = \{C:sm_0(o_i) < Th \}$, $N_{TN} = |N \cap NB|$, $N_{FN} = |N \cap B|$.

Another balanced and integral metric is the Receiver Operating Characteristic (ROC) curve, which is a set of recall and (1-specificity) tuples calculated for different threshold values. This can be viewed as multiple slices of A/B testing: $\forall th_l \in Th=\{th_1 \dots th_L\} \, , \exists roc_l = recall(th_l) \times (1-specificity(th_l))$. 

The Area Under the Curve (AUC) provides an aggregate measure of the ROC curve. We approximate the AUC using the $30$ threshold values in $Th$ for our ROC/AUC calculations:

\begin{equation}
\label{eq:10}
AUC = \sum_{l=1}^{L}(roc_l+roc_{l-1})/(2*((specificity_{l-1}-specificity_{l})))
\end{equation}

For the accuracy metrics presented below—Accuracy, Precision, Recall, Specificity, and F1 score—we set the trusted threshold to a neutral value of 
$Th=0.5$ value. If minimizing false-positive or false-negative errors is more critical, adjusting the $Th$ value is an effective strategy, and the ROC curve is a useful tool for determining this optimal threshold. However, due to publication size constraints, the optimization study for $Th$ is beyond the scope of this paper.

\section{Results}
For the activation function control parameter, only three architectures converged (ReLU, KGate, Tanh), while Linear regression, Sigmoid, RBF, and Transformer-like non-linearity models failed to converge in training, despite parameter tuning. Tanh converged to the accuracy equivalent of the ReLU and KGate (implementation of the GLU class activation function) with about $3$ times larger training iterations.

\begin{table}
\caption{Accuracy metrics and training time for three convergent models.}
\label{tab.models}
\begin{tabular}{|l|c|c|c|c|c|c|c|}
\hline
Model & Precision & Recall & F1 & FN & FP & AUC & Time [s]\\
\hline
ReLU & 0.60135 & 0.87689 & 0.71344 & 449 & 2120 & 0.99687 & 4547\\
KGate & 0.39418 & 0.91774 & 0.55149 & 300 & 5144 & 0.99254 & 4822\\
Tanh & 0.57488 & 0.72416 & 0.64094 & 1006 & 1953 & 0.98401 & 11088\\
\hline
\end{tabular}
\end{table}

The default training parameters included a contact window of the protein residue and RNA base of the same length $15$, which resulted in the unbalanced training set consisting of $196,862$ bind examples and $1,8607,564$ non-bind examples. Because of the hardware, particularly ROM limitations, at any moment of time, a maximum of $9,843,100$ non-bind examples were loaded into memory to be used fully or partially for randomly drawing training subsets from them. The test set has $3,647$ binding and $919,837$ non-binding sequences, and accuracy for all models is around $0.99+$, which remains the same in other experiments unless it is mentioned otherwise. Other accuracy metrics and training time are shown in Table~\ref{tab.models}.

\begin{table}
\caption{Accuracy metrics and training time for ReLU model for re-balancing techniques.}
\label{tab.bal}
\begin{tabular}{|l|c|c|c|c|c|c|c|}
\hline
Re-sampling & Precision & Recall & F1 & FN & FP & AUC & Time [s]\\
\hline
Over & 0.23265 & 0.98876 & 0.37667 & 41 & 11894 & 0.99615 & 3612\\
Under & 0.13755 & 0.99369 & 0.24165 & 23 & 22723 & 0.99077 & 165\\
No & 0.60135 & 0.87689 & 0.71344 & 449 & 2120 & 0.99687 & 4547\\
\hline
\end{tabular}
\end{table}

Results for experiments with re-balancing techniques for ReLU model, with default parameters of the window length of the neighbouring to the binding point residues and bases of $15$, are shown in Table~\ref{tab.bal}. The training set for binding oversampling has $\frac{25}{50}$ ratio of binding to non-binding sequences, for non-binding under-sampling, has $\frac{1}{2}$ ratio, and default $\frac{1}{100}$ ratio.

\begin{table}
\caption{Accuracy metrics and training time for ReLU model with re-training on random subsets.}
\label{tab.retr}
\begin{tabular}{|l|c|c|c|c|c|c|c|c|c|}
\hline
Window size & Retrain & Epoch & Precision & Recall & F1 & FN & FP & AUC & Time [s]\\
\hline
15 & 1 & 50 & 0.60135 & 0.87689 & 0.71344 & 449 & 2120 & 0.99687 & 4547\\
15 & 5 & 10 & 0.58072 & 0.88374 & 0.70088 & 424 & 2327 & 0.99716 & 4335 \\
15 & 10 & 5 & 0.59051 & 0.88018 & 0.70682 & 437 & 2226 & 0.99740 & 5120\\
15 & 20 & 2 & 0.64614 & 0.82561 & 0.72493 & 636 & 1649 & 0.99578 & 5462\\
27 & 1 & 100 & 0.54466 & 0.92624 & 0.68596 & 269 & 2824 & 0.99642 & 13766\\
27 & 20 & 4 & 0.51625 & 0.94927 & 0.66879 & 185 & 3244 & 0.99720 & 15263\\
27 & 40 & 4 & 0.52737 & 0.94845 & 0.67784 & 188 & 3100 & 0.99774 & 30811\\
\hline
\end{tabular}
\end{table}
 
Results for experiments with re-training on random subsets for ReLU model. for the window length of the neighbouring to the binding point residues and bases of $15$ and $27$, are shown in Table~\ref{tab.retr}. The training set has $\frac{1}{80}$ ratio of binding to non-binding sequences, and overall training epochs are $50$ and $300$.

\begin{table}
\caption{Accuracy metrics and training time for ReLU model with various window lengths.}
\label{tab.win_len}
\begin{tabular}{|c|c|c|c|c|c|c|c|c|}
\hline
Window size & Precision & Recall & F1 & FN & FP & AUC & N no-bind & Time [s]\\
\hline
39  & 0.62172 & 0.92926 & 0.74500 & 258 & 2062 & 0.99722 & 378556 & 10573\\
35  & 0.56191 & 0.96189 & 0.70940 & 139 & 2735 & 0.99745 & 386473 & 9392\\
31  & 0.59863 & 0.96106 & 0.73774 & 142 & 2350 & 0.99784 & 393868 & 8040\\
27  & 0.67542 & 0.94544 & 0.78793 & 199 & 1657 & 0.99805 & 380116 & 7008\\
23  & 0.78827 & 0.94324 & 0.85882 & 207 & 924 & 0.99899 & 396438 & 5856\\
19  & 0.69987 & 0.91500 & 0.79311 & 310 & 1431 & 0.99685 & 417053 & 4818\\
15  & 0.73047 & 0.87689 & 0.79701 & 449 & 1180 & 0.99653 & 442148 & 4547\\
11  & 0.61301 & 0.80614 & 0.69644 & 707 & 1856 & 0.99124 & 440511 & 2847\\
7  & 0.60935 & 0.51111 & 0.55592 & 1783 & 1195 & 0.96258 & 548721 & 1909\\
3  & - & 0.0 & - & 3647 & 0 & 0.774843 & 856833 & 1031\\
\hline
\end{tabular}
\end{table}

Results for experiments with various window lengths of the neighbouring to the binding point residues and bases for the ReLU model are shown in Table~\ref{tab.win_len}. The training set has $\frac{1}{100}$ ratio of binding to non-binding sequences.

\begin{table}
\caption{Accuracy metrics and training time for ReLU, KGate, Tanh homogeneous model ensembles.}
\label{tab.homo_ens}
\begin{tabular}{|l|c|c|c|c|c|c|c|}
\hline
\multirow{2}{7em}{Ensemble Type, Size and Voting} & Precision & Recall & F1 & FN & FP & AUC & Time [s]\\
& & & & & & &\\
\hline
ReLU 1 & 0.675416 & 0.945435 & 0.787934 &  199 & 1657 & 0.998049 & 7008\\
ReLU 5 consensus & 0.609330 & 0.938305 & 0.738854 & 225 & 2194 & 0.997637 & 28407\\
ReLU 5 majority & 0.568104 & 0.950370 & 0.711120 & 181 & 2635 & 0.997769 & 28407\\
KGate 1 & 0.527786 & 0.958322 & 0.680689 & 152 & 3127 & 0.995889 & 10260\\
KGate 5 consensus & 0.734260 & 0.927338 & 0.819581 & 265 & 1224 & 0.993780 & 50370\\
KGate 5 majority & 0.528442 & 0.957774 & 0.681096 & 154 & 3117 & 0.996750 & 50370\\
Tanh 1 & 0.574880 & 0.724157 & 0.640942 & 1006 & 1953 & 0.984010 & 11088\\
Tanh 5 consensus & - & 0.0 & - & 3647 & 0 & 0.500137 & 50321\\
Tanh 5 majority & 0.703614, & 0.715383 & 0.709449 & 1038 & 1099 & 0.990411 & 50321\\
Tanh 7 majority & 0.602045 & 0.774883 & 0.677617 & 821 & 1868 & 0.993211 & 78734\\
\hline
\end{tabular}
\end{table}

Results for experiments with ReLU, KGate, Tanh homogeneous model ensembles for the window length of the neighbouring to the binding point residues and bases of $27$ are shown in Table~\ref{tab.homo_ens}. The training set has $\frac{1}{100}$ ratio of binding to non-binding sequences, and overall training epochs are $50$ for ReLU and KGate and $150$ for Tanh.

\begin{table}
\caption{Accuracy metrics and training time for heterogeneous mixes of ReLU, KGate, Tanh models ensembles.}
\label{tab.hetero_ens}
\begin{tabular}{|l|c|c|c|c|c|c|}
\hline
Ensemble Type, Size and Voting & Precision & Recall & F1 & FN & FP & AUC\\
\hline
3 ReLU + 3 KGate consensus & 0.725897 & 0.920757 & 0.811797 & 289 & 1268 & 0.995493\\
\multirow{2}{10em}{2 ReLU + 2 KGate + 2 Tanh consensus} & 0.910063 & 0.396764 & 0.552606 & 2200 & 143 & 0.989013\\
& & & & & &\\
3 KGate + 3 Tanh majority & 0.76993 & 0.74417 & 0.75683 & 933 & 811 & 0.99462\\
3 ReLU + 3 Tanh majority & 0.72293 & 0.75048 & 0.73645 & 910 & 1049 & 0.99493\\
3 ReLU + 3 KGate majority & 0.58805 & 0.94955 & 0.72630 & 184 & 2426 & 0.99794\\
\multirow{2}{10em}{2 ReLU + 2 KGate + 2 Tanh majority} & 0.66097 & 0.93392 & 0.77409 & 241 & 1747 & 0.99825 \\
& & & & & &\\
\hline
\end{tabular}
\end{table}

Results for experiments with heterogeneous mixes of ReLU, KGate, Tanh models ensembles for the window length of the neighbouring to the binding point residues and bases of $27$ are shown in Table~\ref{tab.hetero_ens}. The training set has $\frac{1}{100}$ ratio of binding to non-binding sequences, and overall training epochs is $50$ for ReLU and KGate and $150$ for Tanh.

\begin{table}
\caption{Accuracy metrics for ReLU, KGate, Tanh models with and without Dropout layers.}
\label{tab.drop_out}
\begin{tabular}{|c|c|c|c|c|c|c|}
\hline
Model Type & Precision & Recall & F1 & FN & FP & AUC\\
\hline
ReLU & 0.59586 & 0.95448 & 0.73369 & 166 & 2361 & 0.99730\\
ReLU + DropOut & 0.67542 & 0.94544 & 0.78793 & 199 & 1657 & 0.99805\\
Tanh & 0.51401 & 0.43268 & 0.46985 & 2069 & 1492 & 0.95936\\
Tanh + DropOut & 0.38228 & 0.52180 & 0.44128 & 1744 & 3075 & 0.92127\\
KGate & 0.52779 & 0.95832 & 0.68069 & 152 & 3127 & 0.99589\\
KGate + DropOut & 0.50129 & 0.95613 & 0.65774 & 160 & 3469 & 0.99515\\
\hline
\end{tabular}
\end{table}

Results for experiments with ReLU, KGate, Tanh models with and without Dropout layers for the window length of the neighbouring to the binding point residues and bases of $27$ are shown in Table~\ref{tab.drop_out}. The training set has $\frac{1}{100}$ ratio of binding to non-binding sequences, and overall training epochs are $50$ for ReLU and KGate and $150$ for Tanh.

\begin{table}
\caption{Accuracy metrics for our best-performing models in comparison with other methods, see Kashiwagi \cite{Kashiwagi2021}.}
\label{tab.comp}
\begin{tabular}{|c|c|c|c|c|c|c|c|c|}
\hline
\multirow{2}{3em}{Metric} & \multirow{2}{4em}{\textbf{Our ReLU}} & \multirow{2}{4em}{\textbf{Our KGate}} & Kashiwagi & DB bind1 & KYG & OPRA & BindN+ & Pprint\\
&&&&&&&&\\
\hline
Precision & \textbf{0.79} & 0.73 & 0.66 & 0.69 & 0.38 & 0.50 & 0.54 & 0.42\\
Recall & \textbf{0.94} & 0.93 & 0.69 & 0.05 & 0.60 & 0.33 & 0.73 & 0.82\\
F1 & \textbf{0.86} & 0.82 & 0.66 & 0.09 & 0.47 & 0.40 & 0.62 & 0.56\\
\hline
\end{tabular}
\end{table}

The accuracy metrics of two of our best-performing models (ReLU+Dropout single model with a context window length of $23$, KGate $5$-model ensemble with a context window length of $27$) are compared with results obtained in \cite{Kashiwagi2021} on the same data set, as well as results of other methods cited in the same contribution, also applied to the same data set Table~\ref{tab.comp}.

\section{Discussion}

An unexpected outcome of this study was the ``hard fail" of more than half of the ANN architectures during the training phase, in stark contrast to similar applications in financial series tasks \cite{Selitskiy2022}. Models such as Sigmoid, RBF, and Transformer-like architectures did not converge at all during training, unlike the Tanh model, which merely required more resources and demonstrated poorer accuracy as shown in Table~\ref{tab.models}.

Re-balancing techniques for unbalanced datasets proved effective in reducing false-negative errors but were highly inefficient for false-positive errors, as indicated in Table~\ref{tab.bal}. Due to our focus on balanced accuracy metrics, we prioritized continuous re-training in batches for the significantly unbalanced no-bind data portion, given hardware limitations that prevented single-step processing. However, these experiments did not show reliable improvement, reflecting the well-known ANN ``forgetting" problem. Future studies could explore more granular, overlapping partitioning of no-bind training and test examples to mitigate this issue.

A primary objective was to demonstrate that small but wide ANN models using the neighbouring contexts of protein residues and RNA bases near the bind positions could potentially replace the need for explicit structural and geometric information. This would otherwise require significantly larger and more complex models. The significant influence of context window lengths on false-positive and false-negative errors, as well as high accuracy metrics, validates the effectiveness of this approach for protein-RNA bind prediction, as shown in Table~\ref{tab.win_len}.

The disproportion of false-positive and false-negative errors, expected due to the unbalanced dataset, underscores the need to investigate the benefits of model ensembles. Homogeneous ensembles with consensus voting of KGate models reduced false-positive errors to levels comparable with the better-performing single-model ReLU architecture, as seen in Table~\ref{tab.homo_ens}. This suggests that KGate models learn similar binding patterns but with higher dispersion, which could be reduced by methods other than ensemble approaches.

An interesting observation was made with Tanh model homogeneous ensembles with consensus voting, which found no common true positives across models, indicating that Tanh ANNs learn different binding patterns. Despite lower accuracy, this non-codependent behaviour is noteworthy and warrants further study. Heterogeneous ensembles, particularly those including Tanh models, significantly decreased false-positive errors but at the expense of increased false negatives, highlighting the need for further exploration, as noted in Table~\ref{tab.hetero_ens}.

Experiments with DropOut layers to emulate heterogeneous ensembles within a single ANN produced mixed results, as detailed in Table~\ref{tab.drop_out}. ReLU ANNs benefited from DropOut layers, which were thus included in all presented ReLU models. KGate ANNs showed little change, as their GLU architecture inherently acts like a data-driven dropout. Conversely, Tanh ANNs performed worse with DropOut layers, likely due to excessive stochastic variability.

Finally, we compared the accuracy metrics of our best-performing ANN models with other ML models. Kashiwagi \cite{Kashiwagi2021}, for example, is an SVM-based ML model optimizing binding scores through a linear combination of feature scores for protein residue and RNA base sequences and predicted structural features, as summarized in Table~\ref{tab.comp}.

\section{Conclusions and Future Work}

Our best ANN models, trained and tested on the same data, demonstrated significantly superior performance in balanced accuracy metrics such as Precision, Recall, and F1 score compared to sequence SVM models like BindN+, Pprint, Kashiwagi, or analytical spatial structure-based models like DB\_bind1, OPRA, and KYG. The proposed small-footprint ANN models, which can be run on cost-effective hardware accessible to regular researchers, offer a practical first-line tool for identifying prospective RNA-protein binding complexes. These models can serve as preliminary filters before undertaking more labour-intensive and costly laboratory or advanced AI model verifications. More experiment with similar data sets are planned.

However, the issue of ANN ``forgetting" during training on large, imbalanced datasets, especially changing distribution in time, remains unresolved in our proposed models. The ``mixture of experts"-like approach may be a prospective way of addressing the problem \cite{Selitskiy2024Mem}. Additionally, the integration of secondary or spatial structures, as well as the incorporation of physical relationships using GNN and PINN, \cite{Selitskiy2024GNN}, and generative AI architectures, has not yet been achieved. Traditional dot-product attention mechanisms were also found to be ineffective. Consequently, future research will focus on addressing these areas.

\bibliographystyle{splncs04}
\bibliography{ref}

\end{document}